\documentclass[usenatbib, referee]{biom}
\usepackage{booktabs}
\usepackage{multirow}
\usepackage{footnote}
\usepackage{float}
\usepackage{comment}
\usepackage{amsmath}
\usepackage{amsfonts}
\usepackage{titlesec}
\usepackage{bm}
\usepackage{graphicx}
\usepackage{float}
\usepackage{url}
\usepackage{color}
\usepackage{xargs, xstring}
\usepackage{siunitx}[=v2]
\newcommand{\be}{\begin{eqnarray}}
\newcommand{\ee}{\end{eqnarray}}
\newcommand{\bee}{\begin{eqnarray*}}
\newcommand{\eee}{\end{eqnarray*}}
\newcommand{\bi}{\begin{enumerate}}
\newcommand{\ei}{\end{enumerate}}

\newcommand{\Sfunction}{\mathcal{S}(Z_i, D_i)}

\newcommand{\ci}{\perp\!\!\!\perp}

\usepackage[noabbrev,capitalize]{cleveref}
\crefformat{equation}{(#2#1#3)}

\allowdisplaybreaks

\begin{document}

\title{Principal Stratification with Time-to-Event Outcomes}

\author{Bo Liu$^{1}$,
Lisa Wruck$^{2,3}$, and Fan Li$^{1,*}$\email{fl35@duke.edu} \\
$^{1}$Department of Statistical Science, Duke University, North Carolina, USA \\
$^{2}$Department of Biostatistics and Bioinformatics, Duke University School of Medicine, North Carolina, USA \\
$^{3}$Duke Clinical Research Institute, North Carolina, USA}





\begin{abstract}
Post-randomization events, also known as intercurrent events, such as treatment noncompliance and censoring due to a terminal event, are common in clinical trials. Principal stratification is a framework for causal inference in the presence of intercurrent events. Despite the extensive existing literature, there lacks generally applicable and accessible methods for principal stratification analysis with time-to-event outcomes. In this paper, we specify two causal estimands for time-to-event outcomes in principal stratification. For estimation, we adopt the general strategy of latent mixture modeling and derive the corresponding likelihood function. For computational convenience, we illustrate the general strategy with a mixture of Bayesian parametric Weibull-Cox proportional model for the outcome. We utilize the Stan programming language to obtain automatic posterior sampling of the model parameters via the Hamiltonian Monte Carlo. We provide the analytical forms of the causal estimands as functions of the model parameters and an alternative numerical method when analytical forms are not available. We apply the proposed method to the ADAPTABLE trial to evaluate the causal effect of taking 81 mg versus 325 mg aspirin on the risk of major adverse cardiovascular events.    

\end{abstract}



\maketitle

\section{Introduction}
\label{s:intro}
Randomized controlled trial (RCT) is the gold standard in evaluating efficacy and safety of interventions in medicine. However, post-randomization events, also known as intercurrent events, such as treatment noncompliance, including discontinuation and switching, and censoring due to a terminal event, are prevalent in trials. Such intercurrent events are usually confounded and pose challenges to evaluating comparative effectiveness. Standard intention-to-treat (ITT) analysis ignores intercurrent events and compares outcomes between randomized arms. ITT analysis preserves randomization, but it estimates effectiveness rather than efficacy of the intervention and fails to capture treatment effect heterogeneity. N\"{a}ive methods, such as comparing units per their actual treatment status or discarding subjects with intercurrent events, generally lead to biased causal estimates \citep{Rosenbaum84}. 

In a landmark paper, \cite{Angrist96} proposed the instrumental variable (IV) approach to noncompliance, connecting the structural equation model based IV analysis to the potential outcome framework for causal inference. Here randomization itself is an instrument, because it is unconfounded and usually has no direct effect on the outcome, but it affects the actual treatment received, which in turn affects the outcome. 
The key idea is to classify subjects based on their joint potential compliance status under the treatment and control arms and then estimate causal effects specific to the subpopulation in each stratum, such as the compliers average causal effect (CACE). Under suitable assumptions, Angrist et al. proved that, in the special case of binary IV and treatment, the CACE is equivalent to the two stage least square IV estimand. Similar approach was also independently developed by \cite{baker1994paired}. \cite{FrangakisRubin02} later extended the IV approach to noncompliance to the general framework of principal stratification to handle post-treatment confounding, which is applicable to a wide range of scenarios, including truncation by a terminal event such as death \citep{zhang2003estimation}, surrogate endpoints \citep{gilbert2008evaluating} and selection bias in cluster randomized trials \citep{li2022clarifying}.

Though IV is well known in economics, statistics, and social sciences, principal stratification (PS)--as a generalization of the IV approach--has been seldom applied in medicine. The 2018 ICH E9 guidelines for Statistical Principles for Clinical Trials \citep{ICHE9} advocate to use PS to analyze clinical trials with intercurrent events. This spurred an intense interest in PS in regulatory agencies and industry \citep{lipkovich2022using}. However, there are several barriers to the wide adoption of PS in applications. First, survival or more broadly time-to-event outcomes are prevalent in medical research and requires special handling of censoring, but PS with time-to-event outcomes has been seldom investigated; the few exceptions \citep{baker1998analysis, nie2011inference, yu2015semiparametric,blanco2020bounds,wei2021estimation} all rely on advanced or highly-customized statistical methods. Second, because principal strata are latent, PS analysis inherently involves latent mixture structures. Estimation is usually implemented via either the EM algorithm or Bayesian method, both of which require substantial expertise in statistical methods and programming. Third, lack of software package, compounded with the first two challenges, further limits the method's accessibility to applied researchers.  

In this paper, we provide a generally applicable PS method for time-to-event outcomes that are subject to right-censoring. We specify several causal estimands for time-to-event outcomes (Section \ref{sec:estimands}). For estimation, we adopt the general strategy of latent mixture modeling and derive the corresponding likelihood function. For computational convenience, we illustrate the strategy with a mixture of Bayesian parametric Weibull-Cox proportional hazard model for the outcome (Section \ref{sec:structure}). We utilize the Stan programming language to obtain automatic posterior sampling of the model parameters. We provide the analytical forms of the causal estimands as functions of the model parameters and an alternative numerical method when analytical forms are not available (Section \ref{sec:modelspec}). We conduct simulation studies to examine the performance of the proposed methods (Section \ref{sec:simulation}). We apply the proposed method to the ADAPTABLE trial to evaluate the causal effect of taking 81 mg versus 325 mg aspirin on the risk of major adverse cardiovascular events (Section \ref{sec:application}).

\section{Setup and Estimands} \label{sec:estimands}
To convey the main message, we set up PS with time-to-event outcomes in the classic setting of two-arm randomized experiments with all-or-nothing (i.e. binary) compliance. Note that the methods discussed here can be readily adapted to other PS settings, e.g. multiple treatments and continuous compliance. Consider a sample of units $i\ (i=1,\ldots, N)$ from a target population. For each unit $i$, we observe a set of $p$ baseline covariates, $X_i$, and the randomized assigned treatment status, $Z_i(=z)$, with $z=1$ indicating treatment and $z=0$ control. Each unit has a failure time $T_i$, which is subject to right-censoring at time $C_i$. Therefore, we only observe the potentially censored time, $Y_i=\min(T_i,C_i)$, and the censoring indicator, $\delta_{i}=\textbf{1}\{T_{i}\geq C_{i}\}$. To define causal effects, we introduce the notation of potential outcomes \citep{Neyman23}. Assuming the stable unit treatment value assumption (SUTVA) \citep{Rubin78}, each unit has two potential failure times $\{T_{i}(0),T_{i}(1)\}$ and two potential censoring times $\{C_{i}(0), C_{i}(1)\}$, one under each assignment status. Denote the potential survival distribution under assignment $z$ as $G_z(t)=\Pr(T_i(z)>t)$ for $z=1,0$. The ITT survival probability causal effect (SPCE) is:
\begin{equation}\label{eq:SP}
  \tau^{SPCE}(t)= G_1(t)-G_0(t),  \quad 0\leq t \leq t_{\max},
\end{equation}
where $t_{\max}$ is the maximum follow-up time. This estimand represents the causal risk difference between the two arms. 
A second estimand is the ITT restricted average causal effect  (RACE) in survival time, 
\begin{equation}\label{eq:RACE}
    \tau^{RACE}(t)= E[T_{i}(1)\wedge t]-E[T_{i}(0)\wedge t], \quad 0\leq t \leq t_{\max},
\end{equation}
where $a\wedge b=\min(a,b)$. The RACE compares the mean potential survival times restricted by a given time point of interest $t$.

A common estimand in survival analysis is hazard ratio (HR), popularized by the Cox proportional hazard model \citep{Cox72}. However, directly comparing the hazard between two treatment arms does not lead to a valid causal estimand \citep{hernan2010hazards,martinussen2020subtleties}. Generally, defining causal contrasts of hazards is subtle and identification is not always available. Therefore, in this paper we focus on SPCE and RACE estimands. 


Noncompliance occurs when some units do not take the treatment they are assigned to. Let $D_i$ denote the actual treatment received by unit $i$. Noncompliance implies $Z_i \neq D_i$ for some $i$. Because the actual treatment $D$ occurs after the randomization, it has two potential outcomes, $D_i(0)$ and $D_i(1)$, one of which is observed, denoted by $D_i=D_i(Z_i)$, and the other is missing. A principal stratification is the classification of units according to their joint potential treatment-receipt statuses, known as principal strata, $S_i = (D_i(0), D_i(1))$ \citep{Angrist96, FrangakisRubin02}. With binary assignment and treatment, there are four principal strata: \emph{(i)} $S_i\equiv c=(0,1)$, namely compliers, units who would take treatment if assigned to treatment and would take control if assigned to control; \emph{(ii)} $S_i \equiv n=(0,0)$, namely never-takers, units who would take control regardless of the initial assignment; \emph{(iii)} $S_i \equiv a=(1,1)$, namely always-takers, units who would take treatment regardless of the initial assignment; \emph{(iv)} $S_i \equiv d=(1,0)$, namely defiers, units whose actual treatment status is the opposite to the assignment. \cite{FrangakisRubin02} generalized the noncompliance setting to any events occurred between treatment and outcome, known as intercurrent events in clinical trials.  We will use the above nomenclature of noncompliance throughout this paper but note that these strata are often termed differently according to the specific context.

A key property of the principal stratum $S$ is that, by construction, it is not affected by the initial assignment, and thus is a pre-randomization variable, so we can define causal effects conditioning on $S$, i.e. the principal causal effects (PCE). Therefore, for each of the ITT estimand in \eqref{eq:SP}-\eqref{eq:RACE}, we can define its corresponding PCE. Specifically, the principal survival probability causal effect is 
\begin{equation}  \label{eq:SP-PS}
    \tau_s^{SPCE}(t) =\Pr (T(1)>t \mid S  =s) - \Pr (T(0)>t \mid S  =s)
\end{equation}
for $s\in \{c, n, a, d\}$. The effect for the compliers, $\tau_{c}^{SPCE}(t)$, is called the \emph{complier average causal effect (CACE)}, which is an estimand of primary interest in the noncompliance setting. Similarly we can define the principal RACE for $s \in \{c, n, a, d\}$: 
\begin{equation}\label{eq:RACE-PS}
    \tau_s^{RACE}(t)= E[T_{i}(1)\wedge t \mid S_i=s]-E[T_{i}(0)\wedge t \mid S_i=s], \quad 0\leq t \leq t_{\max}.
\end{equation}
Because $\tau^{SPCE}$ and $\tau^{RACE}$ are additive estimands, it is straightforward to show that the ITT effect is a weighted average of the corresponding PCEs:
\begin{equation}
    \tau(t)=\sum_{s={c, n, a, d}}\tau_s(t) \Pr(S=s).
\end{equation}

\section{Identification and Estimation} \label{sec:estimation}
\subsection{Nonparametric identification}
Due to the fundamental problem of causal inference, only one of the two potential treatment status, namely $D_i(Z_i)$, is observed for each unit, and thus the individual stratum membership $S_i$ is not observed. In fact, without any assumptions, each observed cell of $(Z, D)$ is composed of a mixture of two strata, as listed in Table \ref{tab:PScompose}. For example, the untreated units in the control arm, i.e. $Z_i=0, D_i=1$, can be either compliers or never-takers. We denote the set of principal stratum values $s$ that are compatible with each combination of $\{Z,D\}$ by $\mathcal{S}(Z, D)$. Therefore, additional assumptions are required to identify the PCEs and the main task in estimation is to disentangle the latent mixtures from the observed data. 

\begin{table}
    \centering
    \setlength\tabcolsep{6pt}
    \caption{Composition of principal strata in observed cells of assigned and actual treatment $(Z,D)$}
    \begin{tabular}{ccc}
    \toprule
              &  $D=0$ &$D=1$ \\ 
            \midrule
        $Z=0$ & never-takers,  compliers  & always-takers, defiers\\
        $Z=1$ & never-takers, defiers   & always-takers, compliers\\
        \bottomrule
    \end{tabular}
    
    \label{tab:PScompose}
\end{table}

We impose two assumptions in the noncompliance setting with time-to-event outcomes.
\begin{assumption} \label{as:randomization}
(Randomized assignment). $ \{T_i(0), T_i(1), D_i(0), D_i(1)\} \ci Z_i.$
\end{assumption} 
\begin{assumption}\label{as:censoring}
(Conditional independent censoring).  $T_i(z)\ci C_i(z) \mid \{Z_i, S_i, X_i\}$. 
\end{assumption}
Assumption \ref{as:randomization} holds by design in our setting; it can be relaxed to an unconfoundedness assumption that allows conditioning on covariates: $ \{T_i(0), T_i(1), D_i(0), D_i(1)\} \ci Z_i \mid X_i$. Assumption \ref{as:censoring} states that the censoring process is independent of the potential survival outcomes given covariates and principal strata. This holds, for example, when the failure times are subject only to administrative right censoring.  

For non-censored outcomes, \cite{Angrist96} imposed two additional assumptions.
\begin{assumption}\label{as:monotonicity}
(Monotonicity): $D_i(1)\geq D_i(0)$.
\end{assumption}

\begin{assumption} \label{as:ER}
(Exclusion restriction): for stratum $s$ in which $D(0)=D(1)$, i.e., always-takers and never-takers, $T_i(1)$ and $T_i(0)$ are identically distributed: $\Pr(T_i(1)\mid S_i=s)=\Pr(T_i(0)\mid S_i=s)$.
\end{assumption}
Monotonicity (Assumption \ref{as:monotonicity}) rules out defiers. It is usually interpreted as subjects participating a trial with ``good faith'', not intentionally defying the assignment, and is deemed reasonable in most clinical trials. Under monotonicity, the $(Z=0, D=1)$ cell consists of only always-takers and thus, without censoring, the distribution of the potential outcome $T(0)$ of the always-takers, $\Pr(T(0)\mid S=a)$, is fully identified from data in this cell, and the proportion of the always-takers is nonparametrically identified as $\sum_i 1\{Z_i=0,D_i=1\}/\sum_i 1\{Z_i=0\}$. Similarly, the $(Z=1, D=0)$ cell consists of only never-takers and thus, without censoring, the distribution of the potential outcome $T(1)$ of the never-takers, $\Pr(T(1)\mid S=n)$, is fully identified from data in this cell, and the proportion of the never-takers is nonparametrically identified as $\sum_i 1\{Z_i=1,D_i=0\}/\sum_i 1\{Z_i=1\}$. In contrast, distributions of the potential outcomes of the other combinations of randomized arm and stratum, namely, $\Pr(T(1)\mid S=a)$,  $\Pr(T(0)\mid S=n)$, $\Pr(T(1)\mid S=c)$, $\Pr(T(0)\mid S=c)$, need to be disentangled as mixtures in the other cells. Exclusion restriction (ER) (Assumption \ref{as:ER}) assumes away direct effects from the random assignment to the outcome for always-takers and never-takers. Combining ER and monotonicity, the only principal stratum with non-zero causal effect is the compliers. 

For non-censored outcomes, \cite{Angrist96} showed that under Assumptions \ref{as:randomization}-\ref{as:ER}, the CACE is nonparametrically identifiable and can be estimated via moments. However, such moment estimators are restricted and do not utilize the covariate information, which can potentially improve the efficiency \citep{MealliPacini13, long2013sharpening}.  Moreover, right censoring renders closed-form identification infeasible. Alternatively, a latent mixture model approach can flexibly incorporate covariates and accommodate different types of outcomes \citep{ImbensRubin97, hirano2000assessing,MatteiLiMealli13}. Therefore, below we extend the latent mixture modeling approach to estimate the PCEs with time-to-event outcomes.

\subsection{General structure of model-based estimation} \label{sec:structure}
With a slight abuse of notation, we generically denote a probability density or distribution function by $\Pr(\cdot\mid \cdot)$. For each unit $i$, we observe five random variables $\{X_i, Z_i, D_i, Y_i, \delta_i\}$. We assume the joint distribution of these variables of all units is governed by a generic parameter $\theta$, conditional on which the random variables for each unit are \emph{i.i.d.}. Then we express the likelihood of the observed data, under Assumptions \ref{as:randomization}-\ref{as:censoring}, as: 
\begin{longequation}
\begin{small}
\begin{aligned}
  &\prod_{i=1}^N \Pr(X_i, Z_i, D_i, Y_i, \delta_i) \\
  &= \prod_{i=1}^N\sum_{s\in \mathcal{S}(Z_i,D_i)} \Pr(X_i)\Pr(Z_i \mid X_i)\Pr(S_i = s \mid Z_i, X_i) \Pr(D_i \mid  S_i=s, Z_i, X_i) \Pr(Y_i, \delta_i \mid  D_i, S_i = s, Z_i, X_i) \label{eq:lik_factor}\\
  &\propto \prod_{i=1}^N\sum_{s\in \Sfunction}\Pr(S_i = s \mid Z_i, X_i) \Pr(Y_i,\delta_i \mid  S_i = s, Z_i, X_i) \label{eq:lik-reduce}\\
  &\propto \prod_{i=1}^N\sum_{s\in \Sfunction} \Pr(S_i = s \mid Z_i, X_i) \Pr(T_i\geq Y_i\mid  S_i=s, Z_i, X_i)^{\delta_i}\Pr(T_i = Y_i\mid S_i=s, Z_i,  X_i)^{1-\delta_i}  \label{eq:lik-final}
\end{aligned}
\end{small}
\end{longequation}
Three terms in \eqref{eq:lik_factor} become constant with respect to the causal estimands and thus are absorbed by the proportional sign in the third line: \emph{(i)} $\Pr(X)$, because we condition on the covariates $X$ instead of specifying a distribution for $X$; \emph{(ii)} $\Pr(Z_i \mid X_i)$, because of the randomization; \emph{(iii)} $\Pr(D_i \mid  S_i=s, Z_i, X_i)$, because the values of $S_i$ and $Z_i$ jointly determine $D_i$. 
The proportional sign in \eqref{eq:lik-final} holds because, 
under Assumption \ref{as:censoring}, we have
\begin{longequation}
\begin{aligned}
    \Pr(Y_i,\delta_i = 1 \mid S_i, Z_i, X_i) & = \Pr(T_i \geq Y_i, C_i = Y_i \mid  S_i, Z_i, X_i) \propto \Pr(T_i \geq Y_i \mid S_i, Z_i, X_i),\\
    \Pr(Y_i,\delta_i = 0 \mid S_i, Z_i, X_i) & = \Pr(T_i = Y_i, C_i > Y_i \mid S_i, Z_i, X_i) \propto \Pr(T_i = Y_i \mid S_i, Z_i, X_i).
\end{aligned}
\end{longequation}

Under randomization, or more generally the unconfoundedness assumption, $p(T_i (z) \mid  S_i, X_i; \theta_T)=p(T_i \mid S_i, Z_i, X_i; \theta_T)$, and thus the outcome model is equivalent to a model for the potential outcomes. In summary, for model-based PS analysis under Assumption \ref{as:randomization} and \ref{as:censoring}, we need to specify two models: \emph{(i)} a principal strata model (S-model): $p(S_i \mid Z_i, X_i; \theta_S)$, and \emph{(ii)} an outcome model (T-model): $p(T_i \mid S_i, Z_i, X_i; \theta_T)$. Monotonicity and ER are not required for the model-based inference of PS. But monotonicity reduces the number of strata and ER forces the causal effects to be zero in always-takers and never-takers, and thus both reduce the components in the likelihood function \eqref{eq:lik-final} and help to reduce the variance of the PCE estimates \citep{ImbensRubin97}.  

With the S-model and T-model specified, there are two common approaches to estimate the model parameters and consequently the causal estimands. The first approach is via the EM algorithm \citep{Dempster77} to integrate out the missing principal strata and then obtain the maximum likelihood estimate \citep{zhang2009likelihood} . The second approach is through the Bayesian paradigm where we specify a prior distribution for all the parameters and obtain the posterior distribution of these parameters given the data and the models. We choose the Bayesian approach for two reasons. First, for time-to-event outcomes, the likelihood function is complex due to the censoring, which renders the implementation of an EM algorithm challenging. Second, the Bayesian method enables straightforward uncertainty quantification of not only the model parameters but also their derived quantities, e.g. causal estimands. Below we describe the Bayesian models and posterior computation.

\subsection{Model specification and posterior inference}\label{sec:modelspec}
Because the principal stratum $S$ is a categorical variable, we specify a multinomial logistic regression model with a reference stratum $s_0$ (S-model), which we assume to be the same between the two arms: 
\begin{equation} \label{eq:S-model}
    \log \frac{\Pr(S = s \mid Z, X)}{\Pr(S = s_0 \mid Z, X)} = \eta_s + X'\xi_s.
\end{equation}
This model implies that the probability of each stratum is
\begin{longequation} \label{eq:S-form}
    \Pr(S = s_0 \mid Z, X) = \frac{1}{1 + \sum_{l\neq s_0} \exp(\eta_l + X' \xi_l)},\quad \Pr(S = s\mid Z, X) = \frac{\exp(\eta_s + X'\xi_s)}{1 + \sum_{l\neq s_0} \exp(\eta_l + X' \xi_l)}.
\end{longequation}

For the time-to-event outcome, we impose the most popular Cox proportional hazard model \citep{Cox72} (T-model); we note the following method can be readily modified to alternative T-models like the accelerated failure time (AFT) model \citep{Wei92AFT}. The Cox model assumes the hazard function for stratum $s$ and assignment $z$ to be 
\begin{equation}
    h(t; s, z) = h_0(t; s, z)\exp(X_i'\beta_{s, z}).
\end{equation} 
where $h_0(t; s,z)$ is a baseline hazard function. In the classic Cox model, $h_0(t; s, z)$ is specified nonparametrically. However, in Bayesian inference, a parametric $h_0(t; s, z)$ is usually preferred for computational convenience. The most common parametric choice for  $h_0(t; s, z)$ is the Weibull model \citep{Abrams96}: $h_0(t; s, z) = \exp(\alpha_{s, z})t^{\varphi_{s, z} - 1}$ with $\varphi_{s, z} > 0$, based on which the Cox model becomes 
\begin{equation} \label{eq:Weibull-Cox}
    h(t; s, z) = t^{\varphi_{s, z} - 1}\exp(\alpha_{s, z} + X_i'\beta_{s, z}),\quad \varphi_{s, z} > 0.
\end{equation} 
We allow the model parameters $\alpha, \beta, \varphi$ to differ between strata and arms to provide flexibility. 
The probability density function and the survival function are uniquely determined by the hazard function $h(t;s,z)$.
Specifically, under the Weibull-Cox model \eqref{eq:Weibull-Cox}, we can show that
\begin{small}
\begin{align} \label{eq:T-form}
    \Pr(T_i \geq t \mid S_i = s, Z_i = z, X_i) &= \exp\left(-\int_0^t h(u; s, z)\,\mathrm{d}u\right)  =  \exp\left\{-\frac{1}{\varphi_{s,z}}t^{\varphi_{s,z}}\exp(\alpha_{s,z} + X_i'\beta_{s,z})\right\},\\
    \Pr(T_i = t\mid S_i = s, Z_i = z, X_i) &= h(t; s, z)\exp\left(-\int_0^t h(u; s, z)\,\mathrm{d}u\right) \nonumber \\
    &= t^{\varphi_{s,z} - 1}\exp(\alpha_{s,z} + X_i'\beta_{s,z})\exp\left\{-\frac{1}{\varphi_{s,z}}t^{\varphi_{s,z}}\exp(\alpha_{s,z} + X_i'\beta_{s,z})\right\}.
\end{align}
\end{small}
Then, we can plug $Y_i$ in place of $t$ 
into the above formulae to calculate the T-model component in the observed data likelihood \eqref{eq:lik-final}. 


For Bayesian inference, we need to specify prior distribution of the parameters in the S-model and the T model. We choose standard weakly informative priors: a flat prior for the intercepts, $p(\eta_s) \propto 1$ and $p(\alpha_{s,z})\propto 1$; a Gaussian prior for the coefficients: $p(\xi_s) \sim \mathsf{N}(0, \sigma_\xi)$ and $p(\beta_{s,z})\sim \mathsf{N}(0, \sigma_{\beta})$, with large (e.g. 100) prior variance $\sigma_{\xi}$ and $\sigma_{\beta}$; a flat prior for the log shape parameter in the Weibull model $p(\log \varphi_{s,z})\propto 1$. 

Given  the S-model and T-model and the prior distributions of the model parameters $\theta$, posterior sampling of $\theta$ is traditionally obtained via the Monte Carlo Markov chains (MCMC), \emph{e.g.} a Gibbs sampler \citep{Geman1984Gibbs} or a Metropolis-Hasting \citep{Hastings1970} algorithm. In these methods, 
each unit's latent principal stratum $S_i$ is sampled along with the model parameters. In this paper,  we leverage the Stan programming language \citep{Stan, Rstan} to facilitate automatic sampling, bypassing the analytical derivation of MCMC. Posterior sampling in \textbf{Stan} is implemented via the Hamiltonian Monte Carlo method \citep{Neal2011HamiltonMCMC}, which uses the derivatives of the posterior density function to sample efficiently across the parameter space. We provide the \textbf{Stan} code of our simulations in the Online Supplementary. 



Once the posterior samples of the parameters in model \eqref{eq:S-model} and \eqref{eq:Weibull-Cox} are obtained, we can use them to calculate the posterior distribution of the principal causal estimands (\ref{eq:SP-PS}) - (\ref{eq:RACE-PS}) based on their analytical relationship, which is derived below. Both estimands rely on the stratum-specific survival function, $G(t; s, z, \theta)= \mathrm{Pr}(T(z) > t\mid S = s, \theta)=\Pr(T > t\mid S = s, Z=z, \theta),$  which can be decomposed as follows 
\begin{small}
\begin{align}
    \mathrm{Pr}(T > t\mid S, Z, \theta) 
     &= \frac{\int\mathrm{Pr}(T > t \mid S, Z, X = x, \theta) \mathrm{Pr}(S \mid Z, X = x, \theta) \Pr(Z\mid X=x, \theta) \mathrm{Pr}(X = x)\,\mathrm{d}x}{\int \mathrm{Pr}(S \mid Z, X = x, \theta) \Pr(Z\mid X=x, \theta) \mathrm{Pr}(X = x)\,\mathrm{d}x} \nonumber\\
     &= \frac{\int\mathrm{Pr}(T > t \mid S, Z, X = x, \theta) \mathrm{Pr}(S \mid Z, X = x, \theta) \mathrm{Pr}(X = x)\,\mathrm{d}x}{\int \mathrm{Pr}(S \mid Z, X = x, \theta) \mathrm{Pr}(X = x)\,\mathrm{d}x} 
    \label{eq:SF-form}
\end{align}
\end{small}
Given the randomization or the unconfoundedness assumption, the treatment assignment $\Pr(Z\mid X, \theta)$ is a constant with respect to the outcome and thus drops out from the expression as long as its parameters are \emph{a priori} distinct and independent of the parameters in the S-model and T-model. Denote $A_i(s,z,\theta)=\Pr(S=s\mid Z=z, X_i, \theta)$, the analytical form of which under the S-model \eqref{eq:S-model} is given in \eqref{eq:S-form}; denote $B_i(t;s,z,\theta)=\Pr(T_i>t \mid S=s, Z=z, X_i, \theta)$,  the analytical form of which under the T-model \eqref{eq:Weibull-Cox} is given in \eqref{eq:T-form}. 
Plugging the posterior samples of $\theta$ into the analytical form of $A(s,z,\theta)$ and  $B(t;s,z,\theta)$ in Equation \eqref{eq:SF-form}, we can obtain the posterior distribution of the survival function $G(t; s, z)$ for any stratum $s$ and assignment $z$ at any time $t$: 
\begin{equation} \label{eq:SF-posterior}
    \widehat{G}(t; s, z, \theta) = \frac{\sum_{i=1}^N A_i(s,z,\theta) B_{i}(t; s,z,\theta)}{\sum_{i=1}^N A_i(s,z,\theta)}.
\end{equation} 
Then we can obtain any summary statistic, e.g. posterior mean or credible intervals, from the posterior distribution; for example, averaging over the posterior samples of $\theta$ provides the posterior mean of the survival function $\widehat{G}(t; s, z)$.


The posterior distribution of the PCE on the survival probability and the restricted average causal effect are
\begin{align}
    \widehat{\tau}_s^{SPCE}(t) &= \widehat{G}(t; s, 1) - \widehat{G}(t; s, 0), \label{eq:SP-est}\\
    \widehat{\tau}_s^{RACE}(t) & = \int_0^t \widehat{G}(u; s, 1)\,\mathrm{d} u - \int_0^t \widehat{G}(u; s, 0) \,\mathrm{d} u \label{eq:RACE-est}
\end{align}
respectively.  Define $\tilde{B}_i(t; s, z, \theta) = \int_0^t B_i(u; s, z, \theta)\,\mathrm{d}u$. Then the integral in \eqref{eq:RACE-est} can be expressed as
\begin{align}
    \int_0^t \widehat{G}(u; s, z)\,\mathrm{d}u &= \frac{\sum_{i=1}^N A_i(s,z,\theta) \tilde{B}_i(t; s,z,\theta)}{\sum_{i=1}^N A_i(s,z,\theta)}, \label{eq:integral-analytical}
\end{align}

When the T-model is the Weibull-Cox model \eqref{eq:Weibull-Cox}, $\tilde{B}_i$ can be derived analytically. Specifically, $\tilde{B}_i(t; s, z, \theta)$ is given by
\begin{equation*}
    \tilde{B}_i(t; s, z, \theta) = \left[{\varphi_{s,z}c_{s,z}^{{1}/{\varphi_{s,z}}}}\right]^{-1}\gamma\left(\frac{1}{\varphi}, c_{s,z}t^\varphi\right),
\end{equation*}
with  $c_{s,z} = \frac{1}{\varphi_{s,z}}\exp(\alpha_{s,z} + X_i'\beta_{s,z})$ and $\gamma(\alpha, x) := \int_0^x u^{\alpha - 1} \exp(-u)\,\mathrm{d}u$ being the lower incomplete gamma function, which can be readily calculated as a built-in function in most standard statistical software. Consequently, \eqref{eq:integral-analytical} can be easily evaluated; plugging it into \eqref{eq:SP-est}-\eqref{eq:RACE-est} yields the posterior estimates of the two estimands.

When the T-model is not Weibull-Cox, the analytical forms of $\tilde{B}_i$ is usually intractable. Nonetheless, one can obtain the posterior estimates of the PCE estimands numerically based on the posterior samples of the survival function $\widehat{G}(t; s, z, \theta)$ in \eqref{eq:SF-posterior}.  The basic routine for numerically computing the integral is as follows. Let $\{u_k = (k / K)t: k = 0,\dots, K\}$ be $K+1$ equally spaced points in $[0, t]$. With a sufficiently large $K$, the integral can be approximated by the Trapezoid rule as
\begin{equation} \label{eq:integral-numerical}
    \int_0^t \widehat{G}(u; s, z)\,\mathrm{d}u \approx \frac{t}{K}\sum_{k = 1}^K \left[\frac{1}{2}\widehat{G}(u_{k-1}; s, z) + \frac{1}{2}\widehat{G}(u_k; s, z)\right].
\end{equation}
This can be readily modified to more efficient algorithms such as the Simpson's rule \citep{davis2007methods}. 
Posterior distributions of the causal estimands in \eqref{eq:SP-est} and \eqref{eq:RACE-est} can then be calculated by plugging in the numerical estimate \eqref{eq:integral-numerical}  with the posterior samples of $\theta$.

\section{Simulations} \label{sec:simulation}

We conduct simulation studies to evaluate \emph{(i)} the performance of the proposed methods under a range of common settings, and \emph{(ii)} how the exclusion restriction (ER) effects the inference. Throughout the simulations, we maintain monotonicity and limit to the cases without covariates to focus on the main message. 

We simulate a randomized experiment with $N = 2,000$ units. The treatment assignment $Z_i$ is independently drawn from a $Bernoulli(0.5)$. Each unit's principal stratum $S_i$ is generated from the multinomial S-model \eqref{eq:S-model}, with the never-takers as the reference stratum and $(\eta_c = 0.87,\eta_a = -0.51)$. This leads to the probability of 0.25, 0.6, and 0.15 for never-takers, compliers, and always-takers, respectively. Each unit's actual treatment $D_i$ is determined by $Z_i$ and $S_i$. The true uncensored failure time is generated from the Weibull-Cox model \eqref{eq:Weibull-Cox}, with separate parameters for each of the six combination of stratum and the treatment assignment $(s,z)$ with $s\in \{n, c, a\}$ and $z\in \{0,1\}$. In particular, we simulate two scenarios of the underlying truth: \emph{(i)} ER holds; and \emph{(ii)} ER does not holds. When ER holds, the outcome model is the same between assignment $z$ for always-takers and never-takers, and thus we need to specify four different T-models. In contrast, when ER does not hold, we need to specify six T-models. Table \ref{tab:sim-Tmodel} presents the T-model parameters for these two scenarios. Following the independent censoring assumption (Assumption \ref{as:censoring}), we draw the censoring time $C_i$ independently from an exponential distribution with rate 0.3, leading to an event rate being approximately 35\%. This results in approximately 700 events, similar to the number of events in the ADAPTABLE trial. 

\begin{table}[h]
    \caption{Parameters of the true S-model and T-model in simulations}
    \small
    \centering
    \begin{tabular}{cccccccccc}
    \toprule
     & & \multicolumn{4}{c}{Scenario (i): ER} & \multicolumn{4}{c}{Scenario (ii): no ER} \\
     \cmidrule(lr){3-6}\cmidrule(lr){7-10}
     & & \multicolumn{2}{c}{$\Pr(Y(0)\mid S=s)$} &  \multicolumn{2}{c}{$\Pr(Y(1)\mid S=s)$} & \multicolumn{2}{c}{$\Pr(Y(0)\mid S=s)$} &  \multicolumn{2}{c}{$\Pr(Y(1)\mid S=s)$}  \\ 
     \cmidrule(lr){3-4}\cmidrule(lr){5-6}\cmidrule(lr){7-8}\cmidrule(lr){9-10}
      $s$ & $\Pr(S = s)$ & $\varphi_{s0}$ & $\alpha_{s0}$ & $\varphi_{s1}$ & $\alpha_{s1}$ & $\varphi_{s0}$ & $\alpha_{s0}$ & $\varphi_{s1}$ & $\alpha_{s1}$ \\
      \midrule
      $n$ & 0.25 & 2.0 & -3.0 & 2.0 & -3.0 & 2.0 & -4.0 & 1.5 & -3.0 \\
      $c$ & 0.60 & 1.5 & -2.4 & 1.5 & -1.8 & 1.5 & -2.5 & 1.0 & -1.5 \\
      $a$ & 0.15 & 1.0 & -1.2 & 1.0 & -1.2 & 1.0 & -1.0 & 0.6 & 0.0 \\
      \bottomrule
    \end{tabular}
    
    \label{tab:sim-Tmodel}
\end{table}

\begin{figure}
    \caption{Posterior survival probability curves for each stratum-treatment combination with the true outcome being generated from a Weibull-Cox model with or without ER and fitted with a Weibull-Cox model with or without ER. The dashed line is the true survival curve. The solid line is the posterior mean survival curve with the associated 95\% credible band being the shaded area.}
    \centering
    \includegraphics[width = \linewidth]{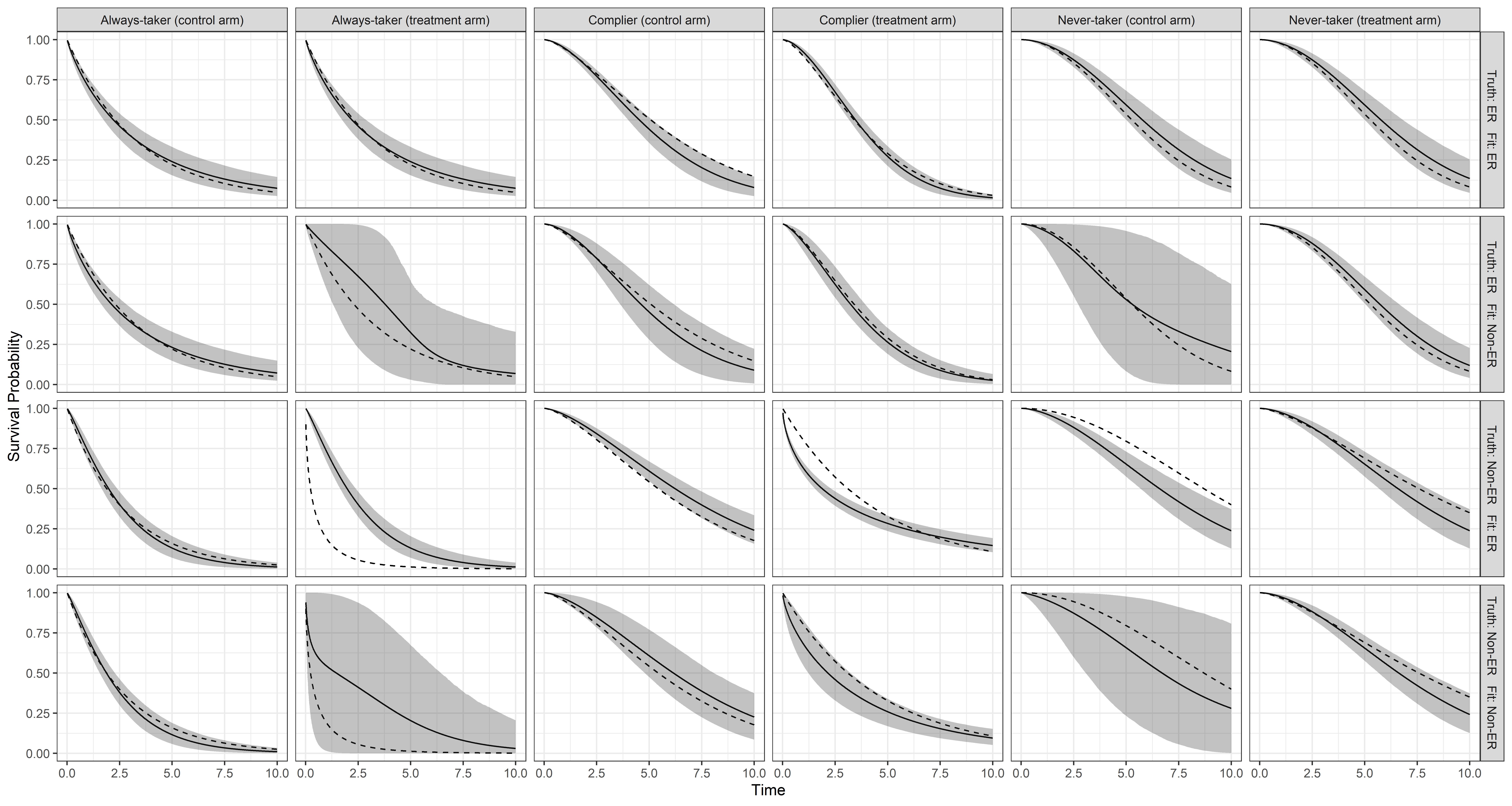}
    
    \label{fig:sim-survival-prob}
\end{figure}

For each simulation scenario, we fit two analyses, with or without ER. So in total we have four combinations of truth and model fit regarding ER. For each combination, we use \textbf{Stan} to run HMC with 6 chains, each chain with 1,000 iterations including 500 warm-up iterations, resulting in a total of 3,000 non-warmup iterations of parameters drawn from the posterior distribution. Mixing of the chains is deemed good under each scenario from the traceplots. The posterior distributions of the survival probability curves under each combination are presented in Figure \ref{fig:sim-survival-prob}. The simulations show that a few patterns. First, in all scenarios, the distributions $\Pr(T(0)\mid S=a)$ and $\Pr(T(1)\mid S=n)$ are correctly estimated with low uncertainty. As discussed before, this is because both distributions are fully identified from the observed data. Second, as shown in the first two panels in Figure \ref{fig:sim-survival-prob}, when the underlying truth satisfies ER, our proposed method recovers the truth regardless of whether assuming ER, but not assuming ER leads to wider credible intervals, particularly for the distributions $\Pr(T(1)\mid S=a)$ and  $\Pr(T(0)\mid S=n)$. This is partially due to the relative low proportion of new-takers and always-takers. Third, when the underlying truth does not satisfy ER but we fit the model with ER, all the four distributions that require disentangling latent mixtures (the middle four graphs in the third panel in Figure \ref{fig:sim-survival-prob}) are estimated with large bias and low uncertainty. In contrast, when the underlying truth does not satisfy ER and we fit the model without ER, the four distributions are covered by their corresponding credible intervals, which are wide but correctly reflect the inherent large uncertainties in disentangling mixtures. Similar patterns are observed in the model parameters and RACE, the results of which are thus delegated to the supporting information (see Web Figure 1, Web Table 1 and Web Table 2). 

\begin{figure}
    \caption{Posterior survival probability curves for each stratum-treatment combination with the true outcome being generated from an AFT model ($\log (T_i) \sim \mathrm{N}(\mu_{sz}, \sigma_{sz})$ for stratum $s$ and arm $z$) without ER, but fitted with a Weibull-Cox model with or without ER. The dashed line is the true survival curve. The solid line is the posterior mean survival curve with the associated 95\% credible band being the shaded area. The parameters $(\mu_{sz}, \sigma_{sz})$ are from left to right, $(1.2, 0.4)$, $(0.8, 0.4)$, $(1.6, 0.2)$, $(1.8, 0.2)$, $(1,0.5)$ and $(1.3, 0.5)$, respectively.}
    \centering
    \includegraphics[width = \linewidth]{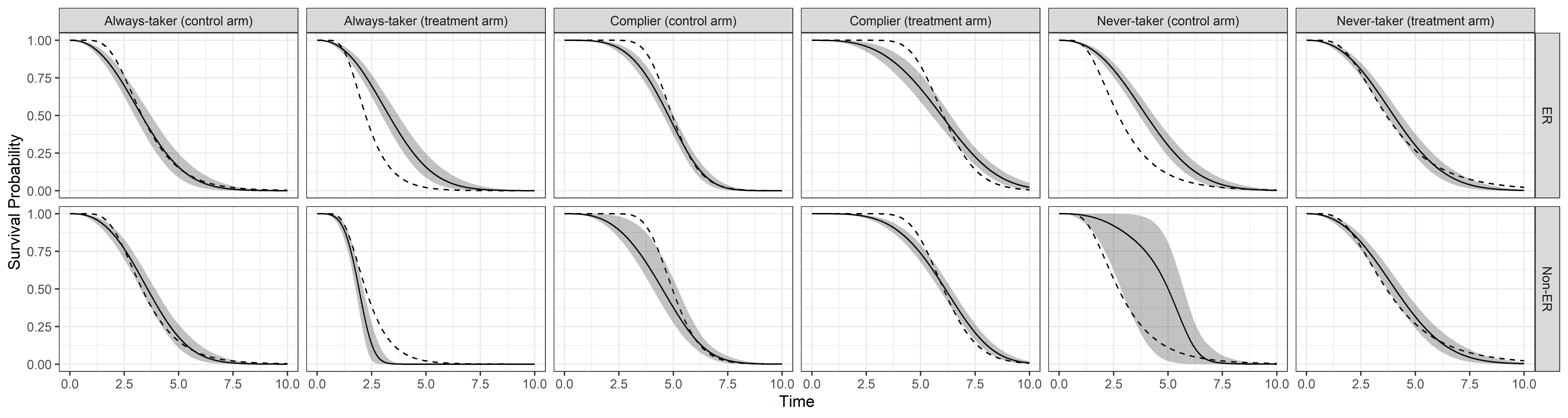}
    \label{fig:sim-survival-prob-AFT}
\end{figure}

In the above simulations the outcome models are correctly specified. Below we consider a more challenging scenario: we simulate the true survival time from an AFT model without ER and with an event rate of about 30\%, but fit the data with the Weibull-Cox model, that is, the outcome model is misspecified in the analysis. The S-model and the censoring model are simulated as before. Figure \ref{fig:sim-survival-prob-AFT} presents the estimated survival probability by stratum and arm. Not surprisingly, the two full nonparametrically identified distributions---$\Pr(T(0)\mid S=a)$ and $\Pr(T(1)\mid S=n)$---are estimated accurately regardless of whether ER being assumed. For the four distributions that requires disentangling mixtures, the estimated survival curves, especially when mistakenly assuming ER, are biased. Nonetheless, when correctly not assuming ER, the bias is relatively small for most cases with the credible bands covering the truth. Estimation in the compliers stratum---the largest stratum---is particularly robust. 

We make several observations from the simulations. First, for the distributions that need disentangling mixtures, the number of units in each stratum directly affects the estimating uncertainties. For example, in our simulations, there are fewest always-takers, followed by new-takers, and most compliers. The discrepancy in the strata size is clearly reflected in the width of the corresponding  credible intervals (Figure \ref{fig:sim-survival-prob}), with  $\Pr(T(1)\mid S=a)$ consistently having the widest interval, followed by $\Pr(T(0)\mid S=n)$, whereas the distribution of the compliers having the tightest intervals. Second, ER plays an important role in the analysis. When correctly assumed, ER can significantly sharpen the inference, but when incorrectly assumed, ER leads to large bias. In real applications, it would be prudent to proceed without ER unless there is strong substantive information suggesting otherwise, and repeat the analysis with ER as a sensitivity analysis. Third, estimation of the nonparametric causal estimands (e.g. survival probability) is largely robust to the specification of the outcome model when ER is not assumed, particularly for strata with large sample sizes, but large uncertainties are common in small strata.

\section{Application to the ADAPTABLE Trial} \label{sec:application}
The ADAPTABLE (\textbf{A}spirin \textbf{D}osing: \textbf{A} \textbf{P}atient-Centric \textbf{T}rial \textbf{A}ssessing \textbf{B}enefits and \textbf{L}ong-Term \textbf{E}ffectiveness) is a multi-center, open label, pragmatic randomized controlled trial designed to assess whether an aspirin intake of 325 mg per day, compared to that of 81 mg per day, would result in a lower risk of major adverse cardiovascular event (MACE), including death from any cause and hospitalization for myocardial infarction, or stroke, with patients with atherosclerotic cardiovascular disease \citep{jones2021comparative}. The trial recruited 15,076 participants, with prior aspirin use reported by 96\% of them, among which the prior regular dose was 81 mg for 85.4\% patients, 162 mg for 2.3\% patients and 325 mg for 12.2\% patients. At the start of the trial, all patients were randomized to 81 mg ($N = 7540$) and 325 mg ($N = 7536$). The primary outcome of the trial is the time to the first occurrence of MACE. The overall rate of MACE during the study is 7.2\%; the median time to event is 26.2 months, with the interquartile range being 19.0 to 34.9 months. During the follow up, 1977 (13.1\%) patients discontinued aspirin, with 15.8\% discontinued in the 325 mg arm and  10.4\% in the 81 mg arm. To illustrate the main method, in this analysis we exclude these patients, but doing so implicitly assumes that the discontinuation behavior is unconfounded, which may not be tenable. More discussion is given in Section \ref{sec:discussion}. Also we restrict the maximum follow-up time 36 months because few events occurred after that. 

Non-adherence to the randomized assignment is prevalent in  ADAPTABLE: 33.5\% of the patients in the 325mg arm switched to low dose, and 6.1\% of the patients in the 81mg arm switched to high dose during the study. The high prevalence of non-adherence in the 325mg arm is partially due to the high dosage often causes gastrointestinal problems among patients. Most of the dose-switching occurred within the first three weeks after randomization and never switched back in the following period; so we treat the dose received at the last visit as a proxy of the actual dose $D_i$, ignoring the exact time when the switch occurs. Table \ref{tab:baseline-characteristics} presents the baseline characteristics by randomized arm and adherence behavior. Almost all covariates are well balanced between the arms, but marked difference in some covariates between adherent and non-adherent patients is observed. For example, there are higher proportion of African Americans, smokers, internet-users, previous GI bleeding among the non-adherent patients in both arms, and generally non-adherent patients tend to switch to their previous aspirin dosage regardless of the randomized dosage. 

\begin{table}[htbp]
\small
\setlength\tabcolsep{1.5pt}
\caption{Baseline characteristics by randomized arm and adherence status in the ADAPTABLE trial.}
    \centering
    \begin{tabular}{lcccccc}
    \toprule
         & \multicolumn{3}{c}{\textbf{Randomized dose: 81 mg}} & \multicolumn{3}{c}{\textbf{Randomized dose: 325 mg}} \\ 
         \cmidrule(lr){2-4} \cmidrule(lr){5-7}
        \textbf{} & \textbf{Overall} & \textbf{Adherent} & \textbf{Non-adherent} & \textbf{Overall} & \textbf{Adherent} & \textbf{Non-adherent} \\
        \textbf{Characteristic} & \textbf{(N=6,556)} & \textbf{(N=6,119)} & \textbf{(N=437)} & \textbf{(N=6,027)} & \textbf{(N=3,698)} & \textbf{(N=2,329)} \\
         \midrule
         Age (year) & 68 (61-74) & 68 (61-74) & 66 (59-73) & 67 (61-73) & 67 (61-73) & 68 (61-74)\\
         Weight (kg) & 91 (79-104) & 91 (79-104) & 92 (80-105) & 91 (78-104) & 92 (80-105) & 89 (77-103)\\
         Female & 1942 (29.6) & 1819 (29.7) & 123 (28.1) & 1856 (30.8) & 1093 (29.6) & 763 (32.8) \\
         Race &\\
         \hspace{1em}White & 5375 (82.0) & 5043 (82.4) & 332 (76.0) & 4979 (82.6) & 3214 (86.9) & 1765 (75.8) \\
         \hspace{1em}Black or African American & 588 (9.0) & 537 (8.8) & 51 (11.7) & 533 (8.8) & 238 (6.4) & 295 (12.7) \\
         \hspace{1em}Other & 361 (5.5) & 333 (5.4) & 28 (6.4) & 294 (4.9) & 153 (4.1) & 141 (6.1) \\
         Hispanic Ethnicity & 211 (3.2) & 194 (3.2) & 17 (3.9) & 176 (2.9) & 96 (2.6) & 80 (3.4) \\
         Current Smoker & 627 (9.6) & 561 (9.2) & 66 (15.1) & 588 (9.8) & 334 (9.0) & 254 (10.9) \\
         Non-internet User & 839 (12.8) & 752 (12.3) & 87 (19.9) & 725 (12.0) & 316 (8.5) & 409 (17.6) \\
         \addlinespace[0.5em]
         \textbf{Medical History} \\
         Coronary Artery Disease & 5990 (91.4) & 5590 (91.4) & 400 (91.5) & 5557 (92.2) & 3395 (91.8) & 2162 (92.8) \\
         Myocardial Infarction & 2337 (35.6) & 2165 (35.4) & 172 (39.4) & 2134 (35.4) & 1266 (34.2) & 868 (37.3) \\
         Coronary-artery Bypass Grafting & 1566 (23.9) & 1464 (23.9) & 102 (23.3) & 1431 (23.7) & 886 (24.0) & 545 (23.4) \\
         Percutaneous Coronary Intervention & 2667 (40.7) & 2494 (40.8) & 173 (39.6) & 2414 (40.1) & 1399 (37.8) & 1015 (43.6) \\
         Cerebrovascular Disease & 1143 (17.4) & 1046 (17.1) & 97 (22.2) & 1026 (17.0) & 587 (15.9) & 439 (18.8) \\
         Hypertension & 5458 (83.3) & 5084 (83.1) & 374 (85.6) & 5041 (83.6) & 3064 (82.9) & 1977 (84.9) \\
         Hyperlipidemia & 5653 (86.2) & 5284 (86.4) & 369 (84.4) & 5232 (86.8) & 3224 (87.2) & 2008 (86.2) \\
         Atrial Fibrillation & 507 (7.7) & 468 (7.6) & 39 (8.9) & 490 (8.1) & 301 (8.1) & 189 (8.1) \\
         Congestive Heart Failure & 1457 (22.2) & 1345 (22.0) & 112 (25.6) & 1415 (23.5) & 814 (22.0) & 601 (25.8) \\
         Peripheral Artery Disease & 1487 (22.7) & 1393 (22.6) & 104 (23.8) & 1432 (23.8) & 803 (21.7) & 629 (27.0) \\
         Diabetes & 2452 (37.4) & 2285 (37.3) & 167 (38.2) & 2303 (38.2) & 1402 (37.9) & 901 (38.7) \\
         Peptic Ulcer Disease & 193 (2.9) & 184 (3.0) & 9 (2.1) & 169 (2.8) & 92 (2.5) & 77 (3.3) \\
         History of Bleeding & 518 (7.9) & 479 (7.8) & 39 (8.9) & 530 (8.8) & 289 (7.8) & 241 (10.3) \\
         Significant Bleeding Disorder & 70 (1.1) & 65 (1.1) & 5 (1.1) & 74 (1.2) & 46 (1.2) & 28 (1.2) \\
         Significant GI Bleed & 391 (6.0) & 369 (6.0) & 22 (5.0) & 389 (6.5) & 210 (5.7) & 179 (7.7) \\
         Intracranial Hemorrhage & 84 (1.3) & 68 (1.1) & 16 (3.7) & 95 (1.6) & 48 (1.3) & 47 (2.0) \\
         \addlinespace[0.5em]
         \textbf{Prior Medications} \\
         Prior Aspirin Use & 6272 (95.7) & 5866 (95.9) & 406 (92.9) & 5765 (95.7) & 3541 (95.8) & 2224 (95.5) \\
         \hspace{1em}Prior Dose: 81 mg & 5320 (84.8) & 5141 (87.6) & 179 (44.1) & 4906 (85.1) & 2910 (82.2) & 1996 (89.7) \\
         \hspace{1em}Prior Dose: 162 mg & 156 (2.5) & 123 (2.1) & 33 (8.1) & 130 (2.2) & 94 (2.7) & 36 (1.6) \\
         \hspace{1em}Prior Dose: 325 mg & 783 (12.5) & 594 (10.1) & 189 (46.6) & 721 (12.5) & 536 (15.1) & 185 (8.3) \\
         P2Y12 Inhibitor & 1420 (21.7) & 1325 (21.7) & 95 (21.7) & 1307 (21.7) & 769 (20.8) & 538 (23.1) \\
         \bottomrule
    \end{tabular}
    
    \label{tab:baseline-characteristics}
\end{table}

\begin{figure}
    \caption{Kaplan-Meier estimate of survival probability for both randomization arms for the first 36 months. The confidence intervals are displayed by the shaded area.}
    \centering
    \includegraphics[width = 0.9\linewidth]{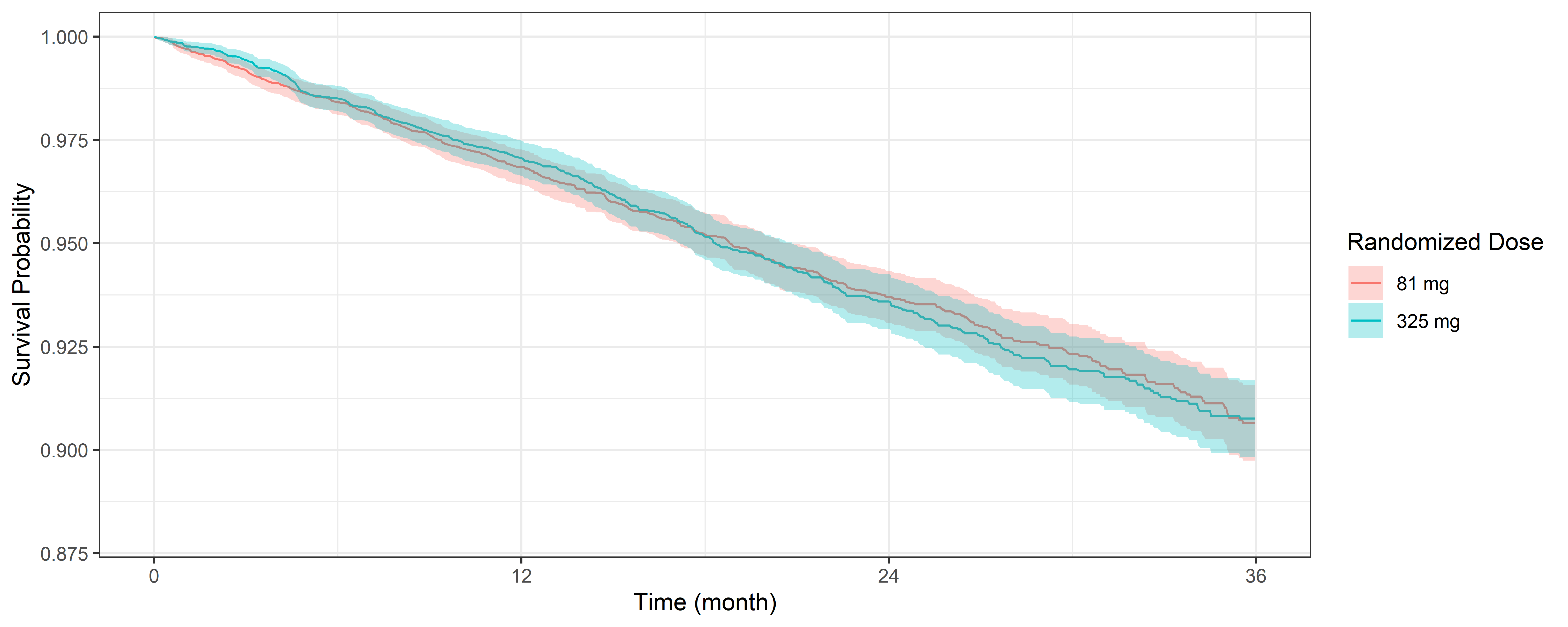}
    
    \label{fig:Kaplan-Meier}
\end{figure}

Figure \ref{fig:Kaplan-Meier} presents the ITT Kaplan-Meier curves of the time to the first occurrence of death or hospitalization by the randomized arms. The two curves are nearly identical throughout the study: the mean time is 26.3 months in each arm, with the standard deviation being 9.71 and 9.67 months for the 81 mg and 325 mg arm, respectively. The event rate is 7.3\% in the 81 mg arm and 7.2\% in the 325 mg arm. The ITT difference in survival probability between the 81 mg and 325 mg arms is 0.002 with 95\% CI (-0.001, 0.014) at 12 months, -0.001 with 95\% CI (-0.019, 0.017) at 24 months, and 0.001 with 95\% CI (-0.025, 0.027) at 36 months. Based on the ITT results, one would conclude that there is no clinical difference in the risk of major adverse cardiovascular event between taking 81 mg versus 325 mg aspirin among patients with atherosclerotic cardiovascular disease \citep{jones2021comparative}. However, the ITT analysis fails to accommodate the nuances of the large proportion of non-adherence and the potential treatment effect heterogeneity. Therefore, we applied the proposed PS method to re-analyze the data.

In the PS analysis, we maintain the randomization, conditional independent censoring and monotonicity assumptions, which are deemed plausible in ADAPTBLE. To reflect the specific context, we term the three strata as always-low ($S = (0, 0)$), consisting of patients who would take 81 mg regardless of the initial randomization, marginal ($S = (0, 1)$), consisting of patients who would take 81 mg if assigned to 81 mg and would take 325 mg if assigned to 325 mg, and always-high ($S = (1, 1)$) patients, consisting of patients who would take 325 mg regardless of the initial randomization. These three strata correspond to never-taker, complier and always-taker, respectively, in the standard nomenclature. Following the remarks from the simulations, we do not impose ER because we cannot rule to direct effects from the randomized assignment to the outcome. 

We fit the multinomial S-model (\ref{eq:S-model}) and the Weibull-Cox T-model (\ref{eq:Weibull-Cox}) to the data. We include covariates that are clinically deemed predictive of the adherence behavior or the outcome. These include patient demographics (age, sex, race, ethnicity), whether the patient is a non-internet user, medical history (atrial fibrillation, percutaneous coronary intervention, bleeding, prior MI) and prior medication use (baseline P2Y12, prior aspirin dose). We use \textbf{Stan} to run 6 chains, each with 1000 iterations including 400 warm-up iterations, resulting in 3,600 posterior draws in total.  The estimated proportion of the always-high, marginal, and always-low stratum is 6.6\% (95\% CI: 6.1\% to 7.2\%), 54.2\%(95\% CI: 52.9\% to 55.5\%) and 39.2\% (95\% CI: 38.0\% to 40.3\%), respectively.  The coefficients of the S-model are delegated to the supporting information (Web Table 3). The coefficients of the T-model are displayed in Table \ref{tab:coef-Tmodel}, showing noticeable heterogeneity between the strata. We provide the stratum-specific summary of selected covariates as follows. In each HMC iteration, we calculate the probability that each patient belongs to each stratum $p_{i,s}$. Then for a given stratum $s$, we calculate the weighted average of the covariates $X_i$ over all patients in the population: $\bar{X}_{s} = {\sum_i p_{i,s} X_i}/{\sum_i p_{i, s}}$, as a summary of the covariates in stratum $s$ in that specific HMC iteration. Repeating this for all iterations provides the posterior distribution of $\bar{X}_{s}$, from which we can calculate the posterior mean and 95\% credible intervals, shown in Table \ref{tab:summary-covariate-stratum}. The strata appear different in many covariates. For example, the marginal patients consist of more white patients and fewer non-internet users. Moreover, patients with history of peripheral artery disease 
are more likely to take low dose regardless of the assignment. Consistent with the previous observation, the most notable difference lies in the prior aspirin dosage: patients tend to take the dose as they had previously taken. 

\begin{table}[ht]
\small
\caption{Fitted coefficients $\beta_{s,z}$ of the Weibull-Cox T-model by stratum $s$ and randomization arm $z$ in the ADAPTABLE trial.}
    \centering
    \begin{tabular}{lcccccc}
    \toprule
    & \multicolumn{2}{c}{Always Low} & \multicolumn{2}{c}{Marginal} &\multicolumn{2}{c}{Always High}  \\
    \cmidrule(lr){2-3} \cmidrule(lr){4-5} \cmidrule(lr){6-7}
                & \multicolumn{1}{c}{81 mg} & \multicolumn{1}{c}{325 mg}  & \multicolumn{1}{c}{81 mg}  & \multicolumn{1}{c}{325 mg}  & \multicolumn{1}{c}{81 mg}  & \multicolumn{1}{c}{325 mg} \\
    \midrule
    \multirow{1}{*}{Age (scaled)} & 0.09 & 0.08 & 0.17 & 0.28 & 0.32 & 0.42 \\
    & (-0.03, 0.22) & (-0.04, 0.20) & (-0.67, 1.18)  & (0.05, 0.53)*  & (0.05, 0.59)* & (-0.54, 1.35) \\
    \multirow{1}{*}{Female} & -0.04 & 0.02 & -0.58 & 0.04 & 0.22 & -0.38 \\
    & (-0.28, 0.20) & (-0.25, 0.28) & (-2.29, 0.93)  & (-0.51, 0.42)  & (-0.36, 0.78) & (-2.19, 1.05) \\
    \multirow{1}{*}{White} & -0.02 & -0.08 & -1.65 & -0.39 & -0.44 & -0.67 \\
    & (-0.28, 0.25) & (-0.37, 0.21) & (-3.10, -0.26)*  & (-0.83, 0.08)  & (-1.01, 0.16) & (-2.29, 0.72) \\
    \multirow{1}{*}{Hispanic} & 0.14 & 0.20 & 1.94 & 0.09 & 0.81 & 1.22 \\
    & (-0.40, 0.27) & (-0.39, 0.75) & (0.51, 3.43)*  & (-0.73, 0.84)  & (-0.03, 1.58) & (-0.15, 2.80) \\
    \multirow{1}{*}{Myocardial Infarction} & 0.31 & 0.41 & -0.90 & 0.33 & -0.07 & -0.45 \\
    & (0.09, 0.53)* & (0.13, 0.67)* & (-2.46, 0.60)  & (-0.06, 0.69) & (-0.60, 0.44)  & (-2.21, 0.97) \\
    \multirow{1}{*}{Atrial Fibrillation} & 0.13 & 0.11 & -0.29 & 0.08 & 0.34 & 0.05 \\
    & (-0.25, 0.49) & (-0.33, 0.49) & (-2.04, 1.31)  & (-0.63, 0.62) & (-0.43, 1.02) & (-1.84, 1.56)  \\
    \multirow{1}{*}{P2Y12} & 0.44 & 0.31 & -0.52 & 0.26 & 0.60 & -0.10 \\
    & (0.21, 0.68)* & (0.04, 0.57)* & (-2.28, 1.08)  & (-0.24, 0.65) & (0.04, 1.13)* & (-1.95, 1.41)  \\
    \bottomrule
    \end{tabular}
    \label{tab:coef-Tmodel}
\end{table}

\begin{table}[ht]
\small
\caption{Stratum-specific summary of pre-treatment variables.}
    \centering
    \begin{tabular}{l
    S[table-format = 2.2]
    S[table-format = 2.2, table-space-text-pre={(}]
    @{,\,}
    S[table-format = 2.2, table-space-text-post={)}]
    S[table-format = 2.2]
    S[table-format = 2.2, table-space-text-pre={(}]
    @{,\,}
    S[table-format = 2.2, table-space-text-post={)}]
    S[table-format = 2.2]
    S[table-format = 2.2, table-space-text-pre={(}]
    @{,\,}
    S[table-format = 2.2, table-space-text-post={)}]
    }
    \toprule
     & \multicolumn{3}{c}{Always-Low} & \multicolumn{3}{c}{Marginal} & \multicolumn{3}{c}{Always-High} \\ \cmidrule(lr){2-4}\cmidrule(lr){5-7}\cmidrule(lr){8-10}
     {Variable} & {Mean} & \multicolumn{2}{c}{95\% CI} & {Mean} & \multicolumn{2}{c}{95\% CI} & {Mean} & \multicolumn{2}{c}{95\% CI} \\ \midrule
        Age (year) & 67.48 & (67.19 & 67.76) & 66.42 & (66.20 & 66.64) & 65.64 & (64.81 & 66.46) \\
        Female (\%) & 32.37 & (31.03 & 33.70) & 28.86 & (27.80 & 29.97) & 30.73 & (26.82 & 34.76) \\
        White (\%) & 75.81 & (74.63 & 76.98) & 87.74 & (86.78 & 88.73) & 77.06 & (73.59 & 80.35) \\
        Hispanic (\%) & 3.81 & (3.29 & 4.35) & 2.59 & (2.14 & 3.02) & 3.78 & (2.31 & 5.46) \\
        Non-internet user (\%) & 18.98 & (17.95 & 19.99) & 7.80 & (6.96 & 8.61) & 20.24 & (16.95 & 23.62) \\
        Myocardial Infarction (\%) & 37.99 & (37.45 & 38.54) & 34.92 & (34.47 & 35.36) & 36.65 & (35.10 & 38.15) \\
        Atrial Fibrillation (\%) & 8.27 & (7.45 & 9.10) & 7.86 & (7.21 & 8.51) & 9.13 & (6.91 & 11.59) \\
        Percutaneous Coronary Intervention (\%) & 44.80 & (43.32 & 46.24) & 38.78 & (37.60 & 39.94) & 39.74 & (35.60 & 43.80) \\
        Bleeding (\%) & 10.19 & (9.42 & 10.97) & 7.21 & (6.59 & 7.86) & 9.75 & (7.44 & 12.37) \\
        P2Y12 (\%) & 24.40 & (23.13 & 25.65) & 21.04 & (20.04 & 22.02) & 22.40 & (18.99 & 26.02) \\
        Prior Aspirin Dose \\
        \hspace{1em} 81 mg (\%) & 87.02 & (86.02 & 88.03) & 83.86 & (82.93 & 84.79) & 43.86 & (39.66 & 48.15) \\
        \hspace{1em} 162 mg (\%) & 1.79 & (1.40 & 2.21) & 2.02 & (1.61 & 2.38) & 7.32 & (5.24 & 9.53) \\
        \hspace{1em} 325 mg (\%) & 8.37 & (7.53 & 9.22) & 10.62 & (9.77 & 11.43) & 43.85 & (39.75 & 47.92) \\
        \bottomrule
    \end{tabular}
    
    \label{tab:summary-covariate-stratum}
\end{table}

\begin{figure}
    \caption{The posterior estimate of survival probability curve for the always-high, marginal and always-low strata under the randomized assignment to 81 mg and 325mg aspirin.} 
    \centering
    \includegraphics[width = \linewidth]{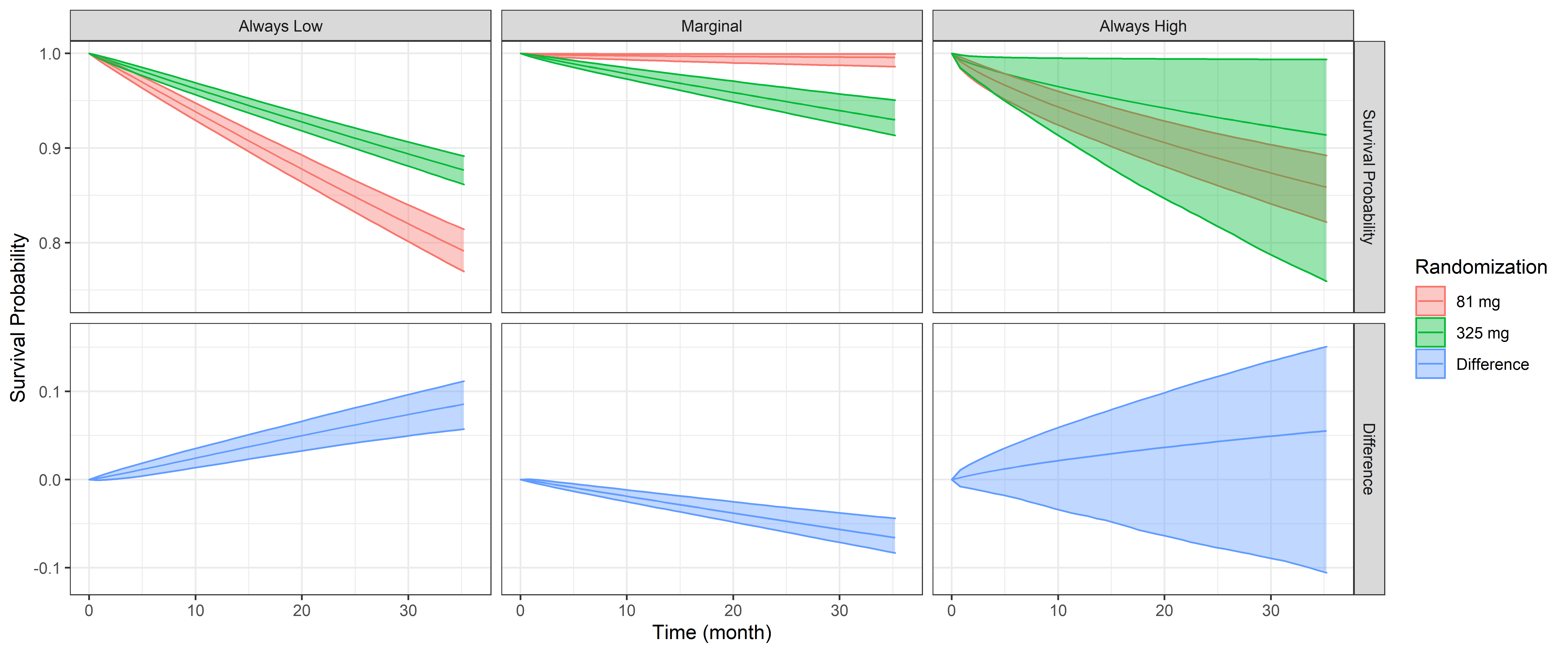}
    
    \label{fig:survival_strata}
\end{figure}

Figure \ref{fig:survival_strata} presents the estimated curves of the potential survival probabilities under each treatment in each stratum. The most striking observation is the heterogeneous treatment effect between the strata. Specifically, for the patients in the marginal stratum, being assigned to 81 mg increases their survival probability comparing with being assigned to 325 mg throughout the follow-up period: the difference in survival probability between the 81 mg and 325 mg
at 12, 24, and 36 months is -0.023 (95\% CI: -0.030 to -0.014), -0.045 (95\% CI: -0.057 to -0.030), and -0.067 (95\% CI: -0.085 to -0.045), respectively. In contrast, for the patients in the always-low stratum, being assigned to 325 mg increases their survival probability comparing with being assigned to 81 mg throughout the follow-up period: the difference in survival probability between the 81 mg and 325 mg at 12, 24, and 36 months is 0.030 (95\% CI: 0.017 to 0.042), 0.060 (95\% CI: 0.040 to 0.078), and 0.087 (95\% CI: 0.058 to 0.114), respectively. The effects are statistically significant as the credible intervals do not cover zero. In comparison, the effects of the assignment for the patients in the always-high stratum are estimated with much larger uncertainties: the difference in survival probability between being assigned to the 81 mg and 325 mg at 12, 24, and 36 months is 0.025 (95\% CI: -0.040 to 0.067), -0.075 (95\% CI: 0.042 to 0.113), and 0.056 (95\% CI: -0.108 to 0.153). The wide intervals are likely due to the low proportion of the always-high stratum (6\%) coupled with the low event rate (around 7\%) in the trial. Similar patterns are observed in the RACE, the details of which are delegated to the supporting information (see Web Figure 2). We conducted a few sensitivity analyses, e.g. fitting an accelerated failure time model, or varying the cutoff time of switching in defining the actual treatment. The results are similar and thus are omitted.  

We shall interpret the clinical meaning of these results with caution. Specifically, for the patients in the always-high and always-low strata, because the randomized assignment does not change their actual treatment, the estimated effects for these two strata are solely attributed to the assignment, similar to a placebo effect. For the patients in the marginal strata, the randomized assignment changes their actual treatment, and thus the effects of the assignment is arguably attributed to the actual treatment. This can be formalized via a separate exclusion restriction assumption for the marginal stratum \citep{ImbensRubinBook}. In summary, comparing with the ITT analysis, the PS analysis of the ADAPTABLE trial provides a more refined picture of heterogeneous treatment effects between subpopulations. In particular, for the patients who would adhere to their assigned dosage regardless of the initial assignment, who comprise over half of the entire patient population, taking 81 mg aspirin appears to consistently and significantly reduce their risk of a major cardiovascular adverse event.

\section{Discussion}
\label{sec:discussion}
We systematically investigated the PS analysis with time-to-event outcomes. We provided a generally applicable method based on the latent mixture models to obtain valid inference of several common causal estimands in the scale of survival probabilities, hazard ratio and restricted mean survival time.  For inference, we adopted the Bayesian approach, leveraging the automatic posterior sampling of the \textbf{Stan} computing platform. Though we focused on the parametric Weibull-Cox model for the outcome, the same inferential framework is applicable to any parametric outcome model specification, such as the AFT model. The Bayesian approach provides a unified framework for inferring all model parameters and derived parameters with automatic uncertainty quantification. It is also straightforward to extend the Bayesian approach to complex cases, such as multiple intermediate variables \citep{mealli2004analyzing},  cluster treatments \citep{frangakis2002clustered}, and quantile effects \citep{wei2021estimation}.

For illustration purpose, in this analysis we excluded the patients population who discontinued the treatment during the study. Because treatment discontinuation is also a post-randomization event that is subject to self-selection, a principled approach would be treating   discontinuation as a separate post-randomization intermediate variable, and jointly analyzing noncompliance and discontinuation using principal stratification \citep{mealli2004analyzing}. However, such an approach, particularly with time-to-event outcomes requires substantial extension in modeling and computation because of the increased number of principal strata, and has remained an open topic.

Besides the mixture modeling approach, another identification and estimation strategy of PS is based on a weighting 
 method under the assumption of \emph{principal ignorability (PI)} \citep{JoStuart09}, which assumes the distributions of the outcomes between strata are the same conditional on the covariates. \cite{jiang2022multiply} proposed a semi-parametrically efficient multiply robust weighting estimator under PI for principal causal effects with non-censored data. That approach is easy to implement and bypass outcome model specification. However, it crucially depends on the PI assumption, the plausibility of which needs to be evaluated on a case by case basis. Also, it is technically non-trivial to extend the weighting method to time-to-event outcomes, which involves counting processes. Nonetheless, such a weighting estimator would be desirable because of its simplicity when the PI assumption is deemed plausible.

\backmatter

\section*{Acknowledgments}
This research is supported by the Patient-Centered Outcomes Research Institute (PCORI) contract ME-2019C1-16146. The contents of this article are solely the responsibility of the authors and do not necessarily represent the view of PCORI. We thank the clinical input and motivating questions from investigators of ADAPTABLE, in particular, Sunil Kripalani, Schuyler Jones, Hillary Mulder.




\bibliographystyle{biom}
\bibliography{PStrata}%

\section*{Supporting Information}

Web Table 1 and Web Figures 1-4 referenced in Section~\ref{sec:simulation} and \ref{sec:application} are available
with this paper at the Biometrics website on Wiley Online Library.

The reproducible \textbf{R} and \textbf{Stan} code in this paper is provided at the Biometrics website on Wiley Online Library and the Github repository: \url{https://github.com/LauBok/PStrata}.

\vspace*{-8pt}

\end{document}